\documentstyle[twoside,fleqn,espcrc2,amssymb,epsfig]{article}

\newcommand{\pspath}{} 
\newcommand{\be}{\begin{equation}}
\newcommand{\ee}{\end{equation}}
\newcommand{\bea}{\begin{eqnarray}}
\newcommand{\eea}{\end{eqnarray}}

\ifx\undefinedcommand\lesssim
\else

\fi

\newcommand{\Reop}{\mathop{\rm Re}\nolimits}

\title{%
  \mbox{}\hbox to 0pt{\vbox to 0pt{\vss
    \parbox[b]{\hsize}{\normalsize
      \begin{flushright}
        HLRZ 55/95\\
        hep-lat/9509042
      \end{flushright}
      \vspace*{2.5cm}}}\hss}%
  Gauge invariant generalization of the 2D chiral Gross-Neveu
  model\footnotemark}

\author{W. Franzki\address{Institute of Theoretical Physics E, RWTH
    Aachen, D-52056 Aachen, Germany}, J. Jers\'ak$^{\rm a}$ and
  R. Welters$^{\rm a}$
  }

\begin{document}

\begin{abstract}
By means of the Lee-Shrock transformation we generalize the 2D Gross-Neveu
(GN$_2$) model to a U(1) gauge theory with charged fermion and scalar
fields in 2D ($\chi U \phi_2$ model).
The $\chi U \phi_2$ model is equivalent to the GN$_2$ model at
infinite gauge coupling.
We show that the dynamical fermion mass generation and
asymptotic freedom in the effective four-fermion coupling persist also when
the gauge coupling decreases. These phenomena are not influenced by the
XY$_2$ model phase transition at weak coupling. This suggests that the
$\chi U \phi_2$ model is in the same universality class as the GN$_2$
model and thus renormalizable.
\end{abstract}

\maketitle

\renewcommand{\thefootnote}{\fnsymbol{footnote}}
\footnotetext{Work supported by DFG and BMBF. The computations have
  been performed on CRAY-YMP of HLRZ J\"ulich.}

\section{Introduction}
Chiral symmetric strongly coupled lattice gauge theories with fermion
and scalar matter fields, $\chi$ and $\phi$, contain unconfined
fermions $F=\phi^\dagger \chi$. Their mass $m_F$ is generated dynamically in
the phase with chiral symmetry breaking \cite{LeShi86d,LeShr87a} and such
models can thus be considered as an alternative to the standard
Higgs-Yukawa mechanism of fermion mass generation, provided they are
nonperturbatively renormalizable \cite{FrJe95a}. Here we consider a
2D model of this type with continuous chiral symmetry, the $\chi U
\phi_2$ model, which in the
strong coupling limit is equivalent to the chiral GN$_2$ model. The
observed scaling properties at large but finite gauge coupling
indicate that the $\chi U \phi_2$ model belongs also here to the
universality class of
the GN$_2$ model. This would mean that the model is renormalizable and
that the shielded gauge mechanism of fermion mass generation, proposed
in \cite{FrJe95a}, works in 2D. A similar problem in 4D is addressed in
ref.~\cite{FrFr95a} and in these proceedings \cite{FrJe96a}.

\section{The model}
The action of the $\chi U \phi_2$ model consists of three parts:
\begin{equation}
S_{\chi U \phi} = S_\chi + S_U + S_\phi
\end{equation}
where
\begin{eqnarray*}
  S_\chi & \hspace{-2mm} = \hspace{-2mm} &
  \frac{1}{2} \sum_x \bar{\chi}_x \sum_{\mu=1}^2
\eta_{\mu x} (U_{x,\mu} \chi_{x+\mu} - U^\dagger_{x-\mu,\mu}
\chi_{x-\mu}) \\
& \hspace{-4mm} & +am_0 \sum_x \bar{\chi}_x \chi_x \;,\\
S_U & \hspace{-2mm} = \hspace{-2mm} &\beta \sum_P (1-\Reop {U_P}) \;,\\
S_\phi & \hspace{-2mm} = \hspace{-2mm} & - \kappa \sum_x
\sum_{\mu=1}^2 (\phi^\dagger_x U_{x,\mu}
\phi_{x+\mu} + H.c.) \;.
\end{eqnarray*}
Here $\chi$ is a staggered fermion field with charge one, $U_{x,\mu}
\in$ U(1) is a compact abelian gauge field with coupling
$\beta=1/a^2g^2$ and $\phi$ is a complex
scalar field with charge one and constraint $|\phi_n|=1$.
The mass term is introduced for technical reasons and the model is
meant in the limit $m_0=0$, where the action has a global U(1)
chiral symmetry.

In the limiting case $\beta=0$ one can perform the
Lee-Shrock-transformation \cite{LeShr87a} in which the scalar and
gauge fields are integrated
out and a four fermion term appears.  This leads to the following action
($r=I_1(2\kappa)/I_0(2\kappa)$):
\begin{eqnarray}
  S_{4f}&=&  - \sum_x \sum_{\mu=1}^2  \Biggl(
  \underbrace{\frac{1-r^2}{4r^2}}_{\displaystyle G}
  \bar{\chi}_x \chi_x \bar{\chi}_{x+\mu} \chi_{x+\mu} \nonumber\\
  &&\hspace{1cm}-\frac{1}{2}\eta_{x,\mu}  \left[ \bar{\chi}_x\chi_{x+\mu}
    - \bar{\chi}_{x+\mu} \chi_x \right] \Biggr) \nonumber\\
  && + \underbrace{\frac{am_0}{r}}_{\displaystyle am'_0} \sum_x
  \bar{\chi}_x\chi_x  \;. \label{wirk}
\end{eqnarray}
\begin{figure}[tbp]
  \begin{center}
    \leavevmode
    \epsfig{file=\pspath 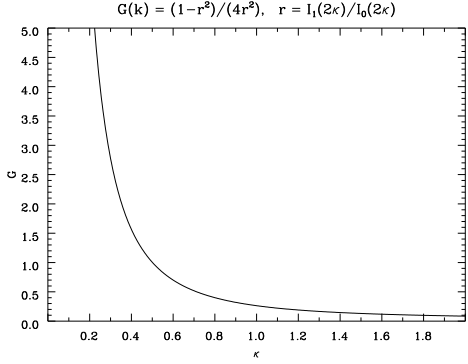,width=5cm,angle=90}
    \vspace{-0.7cm}
    \caption{Relation between the four fermion coupling $G$ and
      $\kappa$ at $\beta=0$.}
    \label{fig:1}
  \end{center}
\end{figure}%
The four fermion coupling $G(\kappa)$ is a function of $\kappa$ which is
shown in fig.~\ref{fig:1}.  The transformed action (\ref{wirk}) for
$am_0=0$ is that of the chiral GN${}_2$ model. This model has a
critical point at $G=0$, where the fermion mass vanishes as
\begin{equation}
am_F \stackrel{G \rightarrow 0}{\propto} e^{\displaystyle
 - \frac{\pi}{8G}}  \;. \label{scal}
\end{equation}

\section{Generalized scaling behavior at $\beta =0$}
We first verified the predicted scaling law (\ref{scal}) at $\beta = 0$.
For $am_0=0$ the simulation is possible only
for small lattices ($ \lesssim  64^2$) because of the
slow convergence of the fermion matrix inversion.

Hence we `somehow' had to extrapolate the masses measured at $am_0>0$
to $am_0=0$. It turned out that all examined schemes which tried to
extrapolate for fixed $\kappa$ to $am_0=0$ failed.
But to perform the continuum limit $\kappa \rightarrow \infty$
it is not necessary to set $am_0=0$ first. Also a combined limit $\kappa
\rightarrow \infty$ and $am_0 \rightarrow 0$ can lead to a continuum theory
without bare mass.

Therefore we looked for a scaling law, in which $am_0>0$ is allowed for
finite $\kappa$ but $am_0 \rightarrow 0$ when $\kappa \rightarrow \infty$.
Such a generalized scaling law is provided by the SD-equations for the
model (\ref{wirk}) truncated to order $O(G)$ \cite{AlGo95}:
\begin{equation}
N=am'_0+ \frac{4G}{V} \sum_k
 \frac{N}{\sum_\nu F_\nu^2 \left(  \frac{1}{a} \sin(k_\nu a)
   \right)^2 + N^2} \label{SDN}
\end{equation}
\begin{equation}
F_\mu = 1+ 2G \frac{a^2}{V} \sum_k \frac{F_\mu \left(  \frac{1}{a}
    \sin(k_\mu a)  \right)^2 } {\sum_\nu  F_\nu^2  \left(  \frac{1}{a}
    \sin(k_\nu a)  \right)^2 + N^2}  \label{SDF}
\end{equation}
These two coupled equations for functions $N$ and $F$,
with $am_F=N/F$, were solved numerically \cite{FrJe96c}.
For infinite volume and small $G$ their approximate analytic solution
is
\begin{eqnarray}
N &=& am'_0 -\frac{8G N}{\pi F^2} \ln \left( \frac{2N}{\pi F} \right)
\label{eq:N} \\
F &=& \frac{1}{2} + \frac{1}{2} \sqrt{1+G} \;.
\end{eqnarray}
For $am_0=0$ one obtains the scaling behavior
(\ref{scal}) as $G \rightarrow 0$.

\begin{figure}[tbp]
  \begin{center}
    \leavevmode
    \epsfig{file=\pspath 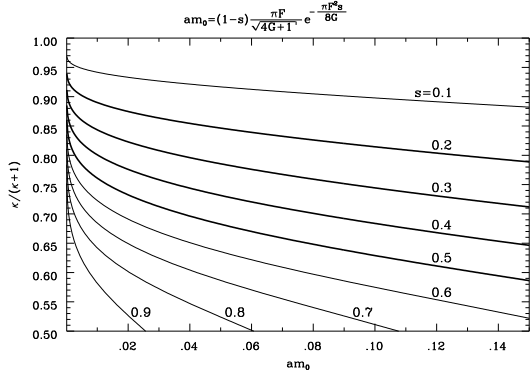,width=5cm,angle=90}
    \vspace{-0.7cm}
    \caption{Approaches to $\kappa=\infty$ and $am_0=0$ for fixed $s$
      according to eq. (\ref{gl5.1}). The measurements have been
      performed along the bold lines.}
    \label{fig:2}
  \end{center}
\end{figure}

The idea of the combined limit $\kappa \rightarrow \infty$ and $am_0(\kappa)
\rightarrow 0$ is to make $am'_0$ a function of $\kappa$ in such a
way, that eq.\ (\ref{eq:N}) is solvable and that $\lim_{\kappa
  \rightarrow \infty}
am'_0(\kappa) =0$.  We choose (see fig.~\ref{fig:2})
\begin{equation}
  am'_0=\frac{1}{r}am_0(G,s) = (1-s)\frac{\pi F}{2}e^{\displaystyle
    -\frac{\pi F^2 s}{8 G}}     \label{gl5.1}
\end{equation}
with a free parameter $s$ obeying $0 < s \le 1$.
The generalized scaling law is then
\begin{equation}
  N=\frac{\pi F}{2} e^{\displaystyle -\frac{\pi F^2 s}{8 G}}
\end{equation}
and thus
\begin{equation}
  am_F=\frac{N}{F}=\frac{\pi}{2} e^{\displaystyle -\frac{\pi F^2 s}{8 G}}
   = \frac{am_0(G,s)}{r F(1-s)} \;.  \label{gl5.3}
\end{equation}
Obviously, the ratio $\frac{m_0}{m_F}$ does not vanish except if $s=1$.

\begin{figure}[tbp]
  \leavevmode
  \hspace*{-0.7cm}
    \epsfig{file=\pspath 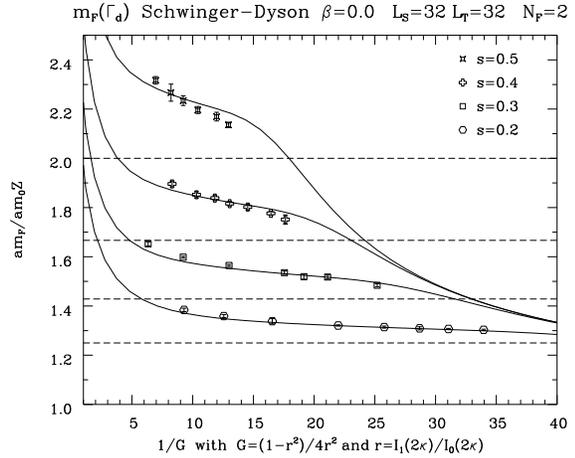,scale=0.32,angle=90}
    \vspace{-1.0cm}
    \caption{Ratios of the physical and bare fermion masses at $\beta
      = 0$ along some lines in fig. \ref{fig:2}. Full lines are
      predictions of the SD-equations for $V<\infty$ and the dashed
      lines represent the expected asymptotic behavior at $V=\infty$
      ($Z=1/rF$).}
    \label{fig:3}
\end{figure}

Fig.~\ref{fig:3} shows, up to the factor $Z=1/rF$ which approaches 1
when $G \rightarrow 0$, the ratio of the measured fermion mass and the
bare mass plotted against $1/G$.
In this ratio the bare part $am_0$ of the fermion mass is `divided
out'.
Calculations have been performed for $s=0.2, 0.3, 0.4, 0.5$. It turned
out to be very difficult to obtain sufficient data for larger values
of $s$, as much larger lattices would be needed.
The full lines show the corresponding ratio calculated
using the SD-equations for finite lattice and those $am_0>0$ given by
(\ref{gl5.1}).  Equation (\ref{gl5.3}) is represented by horizontal dashed
lines.  Good agreement can be observed, especially for not too small $am_0$ and
not too large $G$.  \emph{Thus at $\beta=0$ we have a suitable analytic
  description of the scaling behavior of the measured fermion mass,
  at least in the interval $0.2 \le s \le 0.5$.}

\section{Scaling behavior at $\beta > 0$}
\begin{figure}[tbp]
  \leavevmode
  \hspace*{-0.7cm}
    \epsfig{file=\pspath 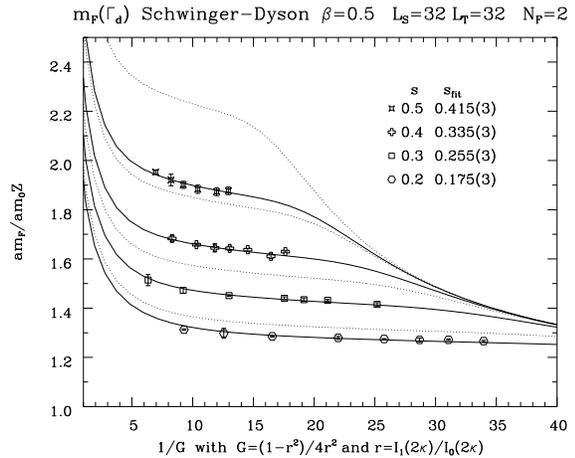,scale=.32,angle=90}
    \vspace{-1.0cm}
    \caption{Data as in fig. \ref{fig:3}, but now at $\beta=0.5$. Full
      lines are solutions of the SD-equations with s replaced by
      $s_{\rm fit}$. Dotted lines show for comparison the same
      solutions at $\beta=0$ as the full lines in fig. \ref{fig:3}.}
    \label{fig:4}
\end{figure}

\emph{But how can the data be described for $\beta \gtrsim 0$?}
Fig.~\ref{fig:4} shows the same ratio, but for masses measured at $\beta=0.5$.
As expected, at the same $G$ value the fermion masses are in general
smaller than those for $\beta=0$, as the gauge coupling becomes
smaller.
This is seen by comparison of the data with the dotted lines which
indicate the ratios calculated by the SD-equations at $\beta=0$.

The data suggest a simple modification of the scaling behavior at
$\beta > 0$ with respect to $\beta =0$. Indeed, we have found that a
good agreement between the measured data and the SD-equations
can be obtained when in these equations the parameter $s$ is replaced
by a fitted value $s_{\rm fit}$.
The full lines in fig.~\ref{fig:4} show the calculated ratio using the
SD-equations (\ref{SDN},\ref{SDF})
with a bare mass which is given by (\ref{gl5.1}) but using $s_{\rm fit}$
instead of $s$.

\emph{Although the theoretical reasons are not well understood,
the SD-equations modified in this way provide a suitable analytic
description of the measured data also for $0< \beta \lesssim 1$ in the
range $0.2 \le s \le 0.5$.}

The scaling law for $\beta \gtrsim 0$ corresponding to
equations (\ref{gl5.3}) and (\ref{gl5.1}) when using $s_{\rm fit}$ is:
\begin{equation}
  m_F=\frac{\pi}{2} D(s,\beta) e^{\displaystyle -\frac{\pi F^2 s}{8
      g_0}} \label{modmF}
\end{equation}
with
\begin{equation}
  D(s,\beta)=\frac{1-s}{1-s_{\rm fit}}   \label{gl5.5}  \;.
\end{equation}
For the examined interval $0.2 \le s \le 0.5$ we thus observe  the
same scaling
law for $\beta \gtrsim 0$ as that for $\beta =0$ apart from the
$G$-independent factor $D(s,\beta)$ with $D(s,0)=1$.

{\em Thus we found some evidence that the $\chi U \phi_2$ model
  for $0 \le \beta \lesssim 1$ and $am_0$ finite but approaching zero
  belongs to the same universality class as the
  GN${}_2$ model with similar bare mass.} A method of extrapolating to
the $am_0=0$ $(s=1)$ limit has
not yet been found, however.

Nevertheless, it is plausible that the scaling behavior remains
consistent with (\ref{modmF}), and thus with the scaling behavior of
the GN$_2$ model, also when $s\rightarrow 1$.

A calculation of $am_F$ for $\beta > 1$ is very difficult because of
its low values, and we could not study its scaling behavior any
more. But looking at some bulk observables and bosonic masses, we
found no indication that the properties of the $\chi U \phi_2$ model
substantially change when $\beta$ is increased. We also found that the
Kosterlitz-Thouless (KT) phase transition occuring at $\beta =
\infty$, when the $\chi U \phi_2$ model reduces to the XY$_2$ model and
decoupled fermions, gets weaker and dissolves at some finite $\beta$
when $\beta$ is increased. This 'remnant' of the KT phase transition does
not influence the fermion mass in an appreciable way. These
observations suggest that the scaling behaviour of $am_F$
is the same at small and large $\beta$, and that only the magnitude of
$am_F$ decreases
with increasing $\beta$. The continuum limit is approached at any
$\beta < \infty$ by turning $\kappa \rightarrow \infty$.

If both above hypothetical extensions of our results are correct, then
the $\chi U \phi_2$ model belongs to the universality class of th
GN$_2$ model at all $\beta < \infty$ and is thus a renormalizable
field theory in which the mass of an unconfined fermion is generated
dynamically. It is thus an example of the shielded gauge mechanisme
proposed in \cite{FrJe95a}.


\bibliographystyle{wunsnot}



\end{document}